\documentclass[aps, pra, twocolumn, amsmath,amssymb,superscriptaddress
]{revtex4-2}

\usepackage{graphicx} 
\usepackage{hyperref} 
\usepackage{mathptmx} 
\usepackage{xcolor}   
\usepackage{physics}  
\usepackage{dcolumn}
\usepackage{tabularx}
\usepackage{orcidlink}

\hypersetup{
    colorlinks=true,
    linkcolor=blue,
    citecolor=blue,
    filecolor=blue,
    urlcolor=blue
}

\newcommand{\UMD}{University of Maryland, College Park, Maryland, 20742, USA}
\newcommand{\SSD}{Sensor Science Division, National Institute of Standards and Technology, Gaithersburg, Maryland 20899, USA}
\newcommand{\CTL}{Electromagnetic Fields Group, National Institute of Standards and Technology, Boulder, Colorado 80305, USA}

\graphicspath{{./}{./figs/}}

\begin{document}

\author{Mingxin Lei\,\orcidlink{0009-0000-6203-2399}}
\affiliation{\UMD}


\author{Stephen P. Eckel\,\orcidlink{0000-0002-8887-0320}}
\affiliation{\SSD}

\author{Eric B. Norrgard\,\orcidlink{0000-0002-8715-4648}}
\affiliation{\SSD}

\author{Nikunjkumar Prajapati\,\orcidlink{0000-0002-7779-9741}}
\affiliation{\CTL}

\author{Alexandra B. Artusio-Glimpse\,\orcidlink{0000-0002-8289-2393}}
\affiliation{\CTL}

\author{Matthew T. Simons\,\orcidlink{0000-0001-9418-7520}}
\affiliation{\CTL}

\author{Christopher L. Holloway\,\orcidlink{0000-0002-4592-9935}}
\affiliation{\CTL}
 
\title{Revisiting collisional broadening of $^{85}$Rb Rydberg levels: conclusions for vapor cell manufacture}

\begin{abstract}
Electrometry based on electromagnetically induced transparency (EIT) in alkali Rydberg vapor cells may suffer reduced sensitivity due to  spurious line broadening effects, caused by surface charges, contaminant gases, or other manufacturing defects.
In order to draw conclusions about the deleterious effects of potential contaminant gases inside Rydberg electrometry vapor cells, we revisit collisional broadening and shifts of both the D$_2$ line and Rydberg levels of rubidium.
Specifically, we measure the broadening and shifts of the $5{\rm S}_{1/2}\rightarrow 5{\rm P}_{3/2}$ (i.e., the D$_2$ line) and $5{\rm S}_{1/2}\rightarrow 5{\rm P}_{3/2}\rightarrow (25{\rm D},27{\rm S},30{\rm D},32{\rm S},35{\rm D},37{\rm S})$ transitions of $^{85}$Rb due to He, Ne, N$_2$ and Ar.
By combining these measurements with observations of velocity changing collisions in the sub-Doppler spectrum of the D$_2$ line, we conclude the following:
(1) that contaminant gases are most likely not the cause of irregular line shapes or shifts of Rydberg transitions due to the high pressures required, and 
(2) the sub-Doppler spectrum of the D$_2$ line, through its accompanying loss of contrast at high pressures, can validate that a vapor cell is sufficiently free of contaminant gas for EIT electrometry.
We use the theory of Omont, {\it J. Phys. France} {\bf 38}, 1343 (1977), to extend our results to a wide variety of possible contaminant gases and further derive scaling laws applicable to all gases.
\end{abstract}

\maketitle

\section{Introduction}
With the advent of Rydberg atom-based RF electrometry, much interest has been given to the need to create high-quality vapor cells that have unperturbed Rydberg spectral lines~\cite{gordon2010,Shaffer20112,6910267,9748947,10238372}.
The high polarizability of Rydberg states enables exquisite RF electric field sensitivity~\cite{gallagher_book}, but also enhances the perturbations due to background electric fields, contaminant gases, and other systematics.
Here we revisit the effect of contaminant gases on the quality of Rydberg spectral lines.
We find:
(1) contaminant gases are most likely not the cause of irregular electromagnetically induced transparency (EIT) line shapes in vapor cells, as roughly 0.02~mbar of contaminant gas would be required to add roughly 1~MHz of additional broadening.
(2) if contaminant gases are to blame, then similar degradation of signal will occur in standard saturation-absorption spectroscopy at similar pressures when using millimeter-sized beams.
We specifically investigated the four background gases N$_2$, He, Ne, and Ar.
We further generalize our results for all possible contaminant gases.

Collisional broadening and shifts of high-lying transitions in alkali atoms have been studied for almost a century.
The first reported measurement by Amaldi and Serg\`{e}~\cite{Amaldi1934} observed the shift in Na $3{\rm S}\rightarrow (8,9,10,11...){\rm P}$ transitions due to H$_2$ gas.
A series of measurements by Fuchtbauer {\it et al.}~\cite{Fuchtbauer1934, Fuchtbauer1935, Fuchtbauer1935a} used a heat pipe, a hydrogen lamp, and a spectrograph to measure the shift and broadening of Na, K, and Cs high-lying P states perturbed by He, Ne, N$_2$, Ar, Kr, Xe, and Hg.
As pointed out by Fermi~\cite{Fermi1934}, the shifts of high-lying Rydberg states should be independent of principal quantum number and even the alkali species.
Rather, the shifts only depend on the s-wave scattering length $a_{\rm S}$ of an electron from the perturber gas and the polarizability of the perturber.
Fuchtbauer applied Fermi's theory to estimate $a_{\rm S}$ for He, Ne, Ar, and N$_2$ and his results were found to be in good agreement with other estimates of $a_{\rm S}$~\cite{massey1969electronic}. 
With the advent of laser spectroscopy, more precise experimental work on pressure shifts of Rb Rydberg atoms includes He, Ne, Kr, Ar, and Xe~\cite{Brillet1980}; He and Ar~\cite{Bruce1982}; He, Ar, and Xe~\cite{Weber1982}; and Ne, Kr, and H$_2$\cite{Thompson1987}.
These works all used direct, two-photon excitation of the excited state.

Likewise, the effects of collisions on the $5{\rm S}_{1/2}\rightarrow 5{\rm P}_{3/2}$ transition (i.e., the D$_2$ line) of Rb has been studied extensively.
Shang-Yi~\cite{Shang-Yi1940} did the first measurements of the broadening and shift of the D$_2$ line due to He, Ar, and H$_2$ gases using a specially constructed vapor cell that could withstand up to 100~bar of the perturber gas.
Absorption line centers, widths, and asymmetries were measured by illuminating the cell with a Ti filament lamp and measuring the absorption through the cell using a grating.
Laser spectroscopy enabled further, higher precision investigations.
Ottinger {\it et al.}\cite{Ottinger1975} studied the effect of He, Ne, Ar, Kr, and Xe on both the D$_1$ and D$_2$ lines.
This study was followed by several more including with Xe, N$_2$ and CH$_4$~\cite{Belov1981}; He, Ne, Ar, Kr, and Xe~\cite{Izotova1981};  He, Ne, Ar, Kr, and Xe looking at the wings of absorption profiles~\cite{Kantor1985}; H$_2$, D$_2$, N$_2$, CH$_4$, and CF$_4$ on both the D$_1$ and D$_2$ lines~\cite{Rotondaro1997}; $^3$He, $^4$He, N$_2$, and Xe on both the D$_1$ and D$_2$ lines~\cite{Romalis1997}; $^3$He and N$_2$ on both the D$_1$ and D$_2$ lines~\cite{Kluttz2013}; and Ne, Xe, N$_2$ on the D$_1$ line~\cite{Ha2021}.
For all of the above studies, pressures of 1~bar to 100~bar were used, and no sub-Doppler features were resolved.
Sargsyan {\it et al.}~\cite{Sargsyan2021} measured collisional shifts due to Ne with EIT sub-Doppler effects.
To the authors' surprise, we could find no studies using the sub-Doppler features of standard saturated absorption spectroscopy~\cite{demtroder2013laser} to measure the collisional broadening and shifts of either the D$_1$ or D$_2$ line.

\begin{figure}
    \center
    \includegraphics{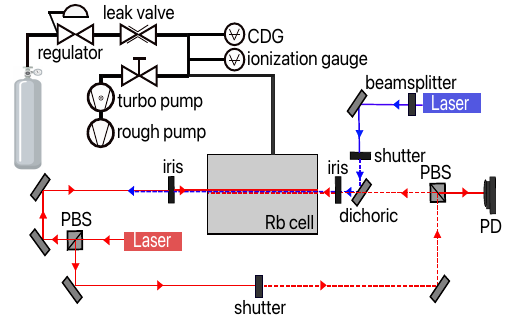}
    \caption{\label{fig:experimental_diagram} Experimental layout of the apparatus.  The gas handling system is in the upper left portion of the diagram, the cell and laser systems are shown in the lower portion. Abrreviations: CDG: capacitance diaphragm gauge, PD: photodiode, PBS: polarizing beam splitter.}
\end{figure}

Here, we fuse together knowledge of D$_2$ collisional broadening and shifts, sub-Doppler spectroscopy, and Rydberg collisional broadening and shifts to draw conclusions for the manufacture and testing of vapor cells for Rydberg electrometry.
We experimentally measure collisional shifts due to He, Ne, N$_2$ and Ar on both the D$_2$ line using sub-Doppler spectroscopy and select Rydberg states of $^{85}$Rb using EIT.
We observe that the effect of velocity changing collisions~\cite{Smith1971, Brechignac1978, Sasso1988, KRONFELDT1994549}, which reduces the contrast of the sub-Doppler Lamb dips, appears at the same order of magnitude of pressure as Rydberg collisional broadening.
Finally, we use theory to expand and scale our results to other gases and justify our conclusions.

\section{Experimental Apparatus}

\begin{figure*}
    \center
    \includegraphics{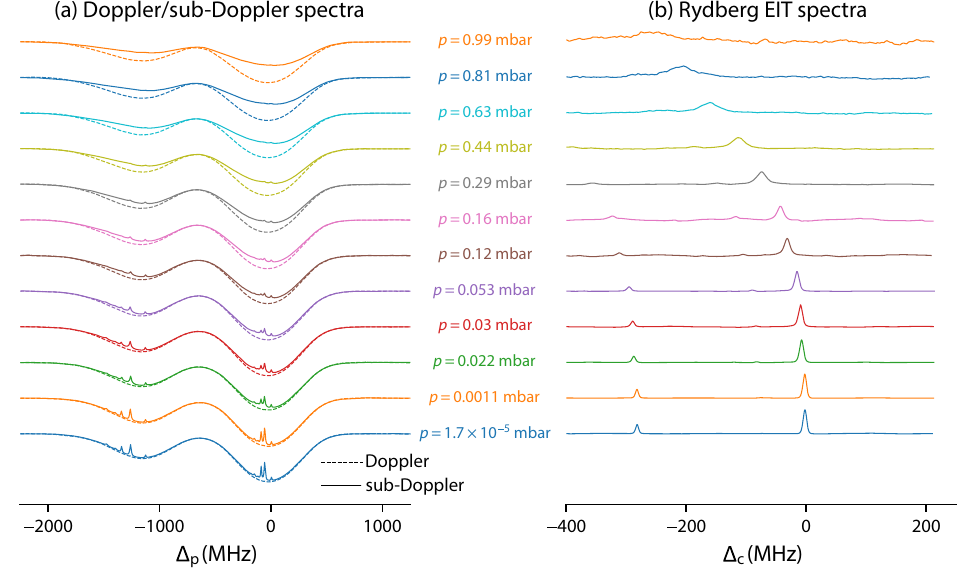}
    \caption{\label{fig:raw_data}
    (Left) Transmission of the probe beam (arbitrary units) with (solid) and without (dashed) the counterpropogating pump beam as a function of probe beam frequency $\Delta_\text{p}$ for different Ar gas pressures.
    Curves at different pressure are marked by different colors and offset for clarity.
    (Right) Transmission of the probe beam with counterpropogating coupling beam as a function of coupling beam detuning $\Delta_{\rm c}$.
    Here, $\Delta_{\rm c}=0$ corresponds to the $^{85}$Rb $5 {\rm P}_{3/2}(F=4)\rightarrow 35 D_{5/2}$ transition.
    Each EIT spectrum is scaled differently to emphasize the relevant features.}
\end{figure*}

Figure~\ref{fig:experimental_diagram} shows a schematic of the experiment.
The experimental cell is composed of a 2.75 inch cube.
An ampule containing roughly 1\,g of Rb was placed in a flexible bellows attached to the cube and broken after the chamber was evacuated.
Two viewports on the cell allow for probe, pump and coupling lasers to pass through the cell along the same axis.
A gas handling system is attached to the cell to allow for the insertion and removal of perturbing gases.

The base pressure of the experimental cell is approximately $2\times10^{-7}$~mbar, consistent with the saturated vapor pressure of Rb at the bellows temperature of 291~K.
We used ultrahigh purity grade (99.999\,\%) of all gases.
The gas pressure was measured by a capacitance diaphragm gauge (CDG) and an ionization gauge.
The relative accuracy of the CDG is specified to 0.15\,\% plus an additional uncertainty of $3\times 10^{-5}$ mbar, although we take the relative uncertainty to be a more conservative 1~\%.
The ionization gauge is calibrated to the CDG for pressures up to a few mbar, and the extrapolated reading is used to record pressures below the usable range of the CDG.

The chamber is surrounded by a polystyrene foam board box to provide thermal insulation.
The box is divided into two sections: one surrounding the cube and one surrounding the bellows.
Each section is monitored by an array of platinum resistance thermometers and temperature stabilized by a forced-air Peltier element.
For these experiments, the temperature of the chamber was stabilized to 304.0(5)~K and the bellows was stabilized at 291.2(2)~K~\footnote{Unless stated otherwise, all uncertainties herein are the combined $k=1$ statistical and systematic uncertainties.}.
With the temperature $T$ and pressure $p$, we use the ideal gas law to predict the density of perturber gas $n = p/(k_BT)$, where $k_B$ is the Boltzmann constant.  
Corrections to the ideal gas law and pressure gradients due to thermal transpiration are negligible at these pressures given our level of uncertainty.

Our experimental sequence is composed of three measurements: Doppler spectroscopy, sub-Doppler spectroscopy, and Rydberg EIT.
In the Doppler spectroscopy sequence, a ``probe'' laser is scanned over the $^{87}$Rb \mbox{$\ket{5S_{1/2},F=2}\rightarrow \ket{5P_{3/2}, F^\prime}$} and $^{85}$Rb $\ket{5S_{1/2},F=3}\rightarrow \ket{5P_{3/2}, F^\prime}$ transitions, a scan of roughly 3.5~GHz width in roughly 50~ms.
We fit the resulting Doppler profile to the model of Ref.~\cite{Siddons2008} to extract the temperature indicated by the Doppler width $T_{\rm D}$, the density of $^{85}$Rb $n_{85}$, and the density of $^{87}$Rb $n_{87}$.
The measured Doppler width corresponds to a temperature of $T_{\rm D}\approx 280$~K, significantly below the measured cell temperature of 304~K.
This discrepancy is due to the probe laser beam power of 30~$\mu$W (saturation parameter $s\approx 1$) being set to optimize the Rydberg EIT signal described below.  This probe power is excessive for accurate Doppler thermometry, as it induces a rather large shift in the measured $T_{\rm D}$ due to hyperfine pumping effects~\cite{Truong2015}.
The fitted densities $n_{85}\approx 8.5\times10^{9}$~cm$^{-3}$ and $n_{87}\approx 3.3\times10^{9}$~cm$^{-3}$ are within an order of magnitude of the predicted saturated density~\cite{Alcock1984} of $n_{85} + n_{87} = 9.8\times10^{10}$~cm$^{-3}$ for $T=291$~K.
 
In the sub-Doppler spectroscopy sequence, an additional counter-propagating 1.5~mW ``pump'' beam with the same frequency to create the well-known sub-Doppler Lamb dips when the both lasers resonantly interrogate the same velocity class.
For both the Doppler and sub-Doppler scans, a frequency ruler is provided by a modulation transfer spectroscopy signal produced in a separate reference cell and the well-known splitting between the $^{87}$Rb $\ket{5S_{1/2},F=2}\rightarrow \ket{5P_{3/2}, F^\prime=3}$  and $^{85}$Rb $\ket{5S_{1/2},F=3}\rightarrow \ket{5P_{3/2}, F^\prime=4}$ transitions~\cite{McCarron2008}.
This allows us to deduce both the shift and broadening of the D$_2$ line as a function of pressure.

For EIT measurements, a counter-propogating, 50~mW ``coupling'' laser is scanned over the transition from $^{85}$Rb $5{\rm P}_{3/2}$ to the target Rydberg level.
During this scan, the the probe laser is locked to the $^{85}$Rb $\ket{5S_{1/2},F=3}\rightarrow \ket{5P_{3/2}, F^\prime = 4}$ transition using the  modulation transfer spectroscopy signal in the reference cell.
For Rydberg S states, the coupling laser is scanned over a range that includes the Doppler crossover resonances with the $\ket{5P_{3/2}, F^\prime = 3,2}$ states.   
For Rydberg D states, the coupling laser is scanned over a range that includes the $D_{3/2}$ and $D_{5/2}$  levels, along with   the Doppler crossover resonances of the respective states.   These known splittings act as a the frequency ruler for the EIT measurements.  
Comparison of the EIT measurements in the experimental cell to a synchronous measurement of EIT in the reference vapor cell determines the frequency shift.

\section{Discussion}

Typical data as a function of pressure is shown in Fig.~\ref{fig:raw_data}, which shows both the Doppler and sub-Doppler spectroscopy sequence and the Rydberg EIT signal perturbed by Ar gas.
At the lowest pressures shown, the large-scale shape of the sub-Doppler and Doppler transmission curves are basically identical, showing the standard Doppler (Gaussian) absorption profile.
The primary difference is Lamb dips that appear in the sub-Doppler spectrum.
As the pressure increases, the sub-Doppler profile and the Doppler profiles begin to diverge, with the Lorentzian peaks becoming smaller and the large-scale Doppler profile of the sub-Doppler Gaussian being reduced. 
For the EIT spectrum, there is a clear shift and broadening as a the pressure increases. 
The pressures at which both effects become apparent is of the order of 0.1~mbar, which is a rather high pressure when one considers that most vapor cells are evacuated to well below $10^{-5}$~mbar before sealing.

Let us now consider the Doppler and sub-Doppler spectra in detail.
The divergence between the sub-Doppler and Doppler signals at $p\gtrsim 0.1$~mbar is caused by velocity changing collisions (VCCs) first recognized in Ne/He mixtures by Smith and H\"ansch~\cite{Smith1971}.
VCCs redistribute the velocities of atoms while they interact with lasers over the whole Doppler velocity distribution; causing the broad Doppler background to reappear in the sub-Doppler spectrum.
Ref.~\cite{Smith1971} employed a hard-collision model, which assumes one collision is sufficient to re-distribute the velocities over the whole of the Doppler profile, to derive a sub-Doppler signal $S_{\rm SD}$ for a two-state system of the form 
\begin{widetext}
\begin{equation}
    S_{\rm SD}(\Delta) \propto \left\{\frac{1}{\Gamma_{\rm vcc}\tau_{\rm R} + 1}\left(\frac{\Gamma}{2\pi}\frac{1}{\Delta^2 + \Gamma^2/4}\right) + \frac{\Gamma_{\rm vcc}\tau_{\rm R}}{\Gamma_{\rm vcc}\tau_{\rm R} + 1}\left[\frac{1}{\sqrt{2\pi (\delta\nu_{\rm D})}} \exp\left(\frac{-\Delta^2}{2(\delta \nu_D)^2}\right)\right]\right\}\exp\left(\frac{-\Delta^2}{2(\delta \nu_D)^2}\right),
    \label{eq:vcc_form}
\end{equation}
\end{widetext}
where $\Gamma_{\rm vcc}$ is the rate of VCCs, $\tau_{\rm R}$ is the relaxation rate of the ground state, $(\delta\nu_D)^2 = k_B T/(2\pi m)$, $m$ is the mass of a rubidium atom, and $\Gamma$ is the homogeneous linewidth measured in Hz, and $\Delta$ is the detuning measured in Hz.
Here, we have ignored cross-relaxation effects in the excited state.
The lineshape \eqref{eq:vcc_form} is a sum of an (area-normalized) Lorentzian and an (area-normalized) Gaussian, and the parameter $\Gamma_{\rm vcc}\tau_{\rm R}$ smoothly transitions between the two.

We fit for the sub-Doppler component of the spectrum by first subtracting the sub-Doppler spectrum from the Doppler spectrum, accounting for approximately 2~\% differences in probe laser intensity and photodiode signal baseline between the two scans.
We then fit the sub-Doppler spectrum to a sum of \eqref{eq:vcc_form} for all six Lamb dips.
In total, there are 18 fit parameters per spectrum: 12 amplitudes for all the peaks, three parameters to parameterize the frequency axis shift and non-linearity, $T$, $\Gamma$, and the product $\Gamma_{\rm vcc}\tau_{\rm R}$. 

\begin{figure}
    \center
    \includegraphics{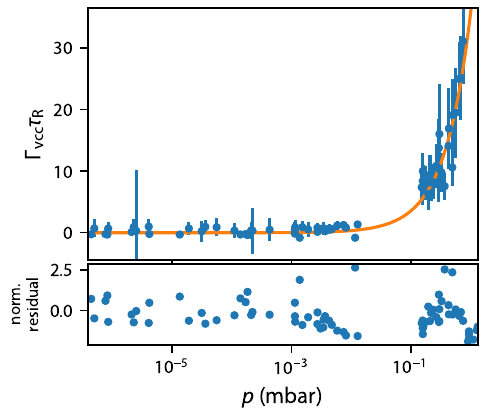}
    \caption{\label{fig:vcc_vs_p} (Top) Dependence of the fit parameter $\Gamma_{\rm vcc}\tau_{\rm R}$ as a function of He gas pressure $p$.  Blue points shown the data, orange curve the linear fit. (Bottom) Normalized residuals of the fit.}
\end{figure}

The fitted parameter $\Gamma_{\rm vcc}\tau_{\rm R}$ follows a linear relationship with pressure, as shown in Fig.~\ref{fig:vcc_vs_p}.
Indeed, one expects that $\Gamma_{\rm vcc} = \langle \sigma v \rangle n$, where $\sigma$ is the cross section for collision with the perturber atom or molecules, $v$ is their relative velocity, $\langle \cdots \rangle$ denotes thermal averaging, and $n$ denotes the density of the perturber gas.
Semi-classically, $\langle \sigma \bar{v}\rangle$ has been derived in the context of the cold atom vacuum standard~\cite{Eckel_2018, Barker2022, Klos2023} and is
\begin{equation}
    \label{eq:thr_vcc_rate}
    \langle \sigma v\rangle = 6.437\ 73\times\left(\frac{k_B T}{E_6}\right)^{3/10} \frac{\hbar}{\mu}x_6\,
\end{equation}
where $E_6 = \hbar^2/2\mu x_6^2$, $x_6=(2\mu C_6/\hbar^2)^{1/4}$, $\mu$ is the reduced mass of the collision pair, and $C_6$ is the Van der Waals coefficient.
Table~\ref{tab:k_vcc} shows estimates of $\langle \sigma v\rangle$ using \eqref{eq:thr_vcc_rate} and $C_6$ coefficients from Klos {\it et al.}~\cite{Klos2023}.

We use our fitted slope for N$_2$ and the semiclassical estimate shown in Table~\ref{tab:k_vcc} to extract $\tau_{\rm R}=0.9$~$\mu$s.
This value of $\tau_{\rm R}$ is significantly smaller than the expected value due to transit broadening of $\tau_{\rm R}\approx 6(2)$~$\mu$s.
Moreover, it is smaller than $\tau_{\rm R}$ values measured directly in similar experiments by modulating the pump beam at different frequencies~\cite{KRONFELDT1994549, THORNTON20112890}.
It is unclear what might be causing the discrepancy in expected $\tau_{\rm R}$; however, it is unimportant for the present discussion.

\begin{table}
    \centering
    \begin{tabular}{D{.}{.}{4.5}D{.}{.}{4.5}D{.}{.}{4.5}D{.}{.}{4.5}c}
         \multicolumn{1}{c}{He} &  \multicolumn{1}{c}{Ne} &  \multicolumn{1}{c}{N$_2$} &  \multicolumn{1}{c}{Ar} & \\
         \hline
        1.60(11) & 2.47(17) & 3.5 & 2.92(16) & This work \\
        \multicolumn{1}{c}{---} &  \multicolumn{1}{c}{---} &  \multicolumn{1}{c}{---} & 0.65 & \cite{THORNTON20112890} \\
        \hline
        2.5 & 2.2 & 3.5 & 3.2 & Eq.~\eqref{eq:thr_vcc_rate} \\
        \hline\hline
    \end{tabular}
    \caption{Estimates of the velocity changing collision rate coefficient $\langle \sigma v\rangle$ in units of $10^{-9}$~cm$^3$/s.
    No uncertainty is specified for our value of N$_2$ because that value is used in combination with \eqref{eq:thr_vcc_rate} to estimate the value of $\tau_{\rm R}$.}
    \label{tab:k_vcc}
\end{table}

Our resulting values of $\langle \sigma v\rangle$ for He, Ne, and Ar are shown in Table~\ref{tab:k_vcc}.
Of these three, Ne and Ar agree with the semi-classical prediction within their uncertainties, whereas He does not.
Our value for Rb-Ar does not agree with the only other measurement of $\langle \sigma v\rangle$, Ref.~\cite{THORNTON20112890}. 
This discrepancy is due to the wildly different estimates of $\tau_{\rm R}$.

\begin{table*}
\begin{tabular}{cD{.}{.}{4.5}D{.}{.}{4.5}D{.}{.}{4.5}D{.}{.}{4.5}D{.}{.}{4.5}D{.}{.}{4.5}D{.}{.}{4.5}D{.}{.}{4.5}c}
     & \multicolumn{2}{c}{$^4$He} & \multicolumn{2}{c}{Ne} & \multicolumn{2}{c}{N$_2$} & \multicolumn{2}{c}{Ar} \\
     Transition & 
     \multicolumn{1}{c}{Shift} & \multicolumn{1}{c}{Broadening} & 
     \multicolumn{1}{c}{Shift} & \multicolumn{1}{c}{Broadening} & 
     \multicolumn{1}{c}{Shift} & \multicolumn{1}{c}{Broadening} & 
     \multicolumn{1}{c}{Shift} & \multicolumn{1}{c}{Broadening} & Ref. \\
    \hline
   ${\rm 5S}\rightarrow{\rm 5P}_{3/2}$ (D$_2$) 
   & -0.22(32) & 6.59(13) & -0.87(12) & 2.57(6) & -2.57(27) & 5.97(18) & -2.32(31) & 6.13(30) & This work \\
   & \multicolumn{1}{c}{---} & \multicolumn{1}{c}{---} & -0.86(8) & 4.1(4) & \multicolumn{1}{c}{---} & \multicolumn{1}{c}{---} & \multicolumn{1}{c}{---} & \multicolumn{1}{c}{---} & \cite{Sargsyan2021} \\
    & \multicolumn{1}{c}{---} & \multicolumn{1}{c}{---} & \multicolumn{1}{c}{---} & \multicolumn{1}{c}{---} & -2.12(5) & 7.01(5) & \multicolumn{1}{c}{---} & \multicolumn{1}{c}{---} & \cite{Kluttz2013} \\
   & 0.151(24) & 7.71(8) & -0.995(8) & 3.86(4) & -2.36(4) & 7.47(16) & -2.35(16) & 7.22(8) & \cite{Rotondaro1997} \\
   & 0.17(2) & 6.74(7) & \multicolumn{1}{c}{---} &  \multicolumn{1}{c}{---} & -2.20(4) & 6.7(1) &  \multicolumn{1}{c}{---} &  \multicolumn{1}{c}{---}  & \cite{Romalis1997} \\
   & 0.67(15) & 6.83(33) & -1.12(15) & 2.85(12) & -2.49(18)
   & 5.73(15) & -2.15(24) & 6.01(15) & \cite{Belov1981, Kantor1985} \\
   & 0.30(7) & 5.0(7) & -0.89(7) & 4.3(1.0) & \multicolumn{1}{c}{---}
   & \multicolumn{1}{c}{---} & -1.89(7) & 6.0(7) & \cite{Izotova1981} \\
   & -0.25(10) & 6.1(1.1) &  -0.66(10) & 3.2(6) &  \multicolumn{1}{c}{---} &  \multicolumn{1}{c}{---} & -2.7(4) & 6.1(8) & \cite{Ottinger1975} \\
   & 0.67 & 7.2 &  \multicolumn{1}{c}{---} &  \multicolumn{1}{c}{---} &  \multicolumn{1}{c}{---} &  \multicolumn{1}{c}{---} &  \multicolumn{1}{c}{---} &  \multicolumn{1}{c}{---}  & \cite{Shang-Yi1940} \\
   \hline
${\rm 5S}\rightarrow {\rm 5P} \rightarrow {\rm 25D}$ & 64.8(7) & 8.2(6) & 5.64(6) & 6.4(4) & 4.2(4) & 24.0(7) & -103.6(1.1) & 21.6(1.2) &  This work \\ 
${\rm 5S}\rightarrow {\rm 25D}$ (2$\gamma$) & \multicolumn{1}{c}{---} & \multicolumn{1}{c}{---} & 6.5(4) & 9.5(6) & \multicolumn{1}{c}{---} & \multicolumn{1}{c}{---} & \multicolumn{1}{c}{---} & \multicolumn{1}{c}{---} & \cite{Thompson1987} \\
  & 57.2(1.6) & 13.4(4) & \multicolumn{1}{c}{---} & \multicolumn{1}{c}{---} & \multicolumn{1}{c}{---} & \multicolumn{1}{c}{---} & -100.4(2.8) & 30.8(2.8) & \cite{Weber1982} \\
  & 77.0(5.0) & \multicolumn{1}{c}{---} & \multicolumn{1}{c}{---} & \multicolumn{1}{c}{---} & \multicolumn{1}{c}{---} & \multicolumn{1}{c}{---} & 131.(13.) & 25.(10.) & \cite{Bruce1982} \\
\hline
${\rm 5S}\rightarrow {\rm 5P} \rightarrow {\rm 27S}$ & 62.1(7) & 10.5(6) & 5.58(17) & 7.4(4) & 4.36(34) & 28.3(1.0) & -106.1(1.2) & 22.7(9) &  This work \\ 
${\rm 5S}\rightarrow {\rm 27S}$ (2$\gamma$) & 54.0(2.0) & 15.0(8) & \multicolumn{1}{c}{---} & \multicolumn{1}{c}{---} & \multicolumn{1}{c}{---} & \multicolumn{1}{c}{---} & -100.(1.6) & 35.4(3) & \cite{Weber1982} \\
\hline
${\rm 5S}\rightarrow {\rm 5P} \rightarrow {\rm 30D}$ & 65.2(7) & 8.7(8) & 5.41(7) & 6.9(4) & 4.91(15) & 23.2(5) & -105.6(1.1) & 17.4(1.1) &  This work \\ 
${\rm 5S}\rightarrow {\rm 30D}$ (2$\gamma$) & \multicolumn{1}{c}{---} & \multicolumn{1}{c}{---} & 6.0(7) & 8.8(8) & \multicolumn{1}{c}{---} & \multicolumn{1}{c}{---} & \multicolumn{1}{c}{---} & \multicolumn{1}{c}{---} & \cite{Thompson1987} \\
  & 58.4(2.0) & 13.6(2.4) & \multicolumn{1}{c}{---} & \multicolumn{1}{c}{---} & \multicolumn{1}{c}{---} & \multicolumn{1}{c}{---} & -102.(4) & 29.6(4.8) &\cite{Weber1982} \\
  & 78.0(4.0) & \multicolumn{1}{c}{---} & \multicolumn{1}{c}{---} & \multicolumn{1}{c}{---} & \multicolumn{1}{c}{---} & \multicolumn{1}{c}{---} & -120.(12.) & 20.(7.) & \cite{Bruce1982} \\
\hline
${\rm 5S}\rightarrow {\rm 5P} \rightarrow {\rm 32S}$ & 65.2(7) & 9.2(1.0) & 5.47(13) & 6.8(1.1) & 5.18(17) & 23.8(8) & -106.8(1.2) & 17.9(7) &  This work \\ 
${\rm 5S}\rightarrow {\rm 32S}$ (2$\gamma$)
 & 56.8(2.4) & 14.0(1.0) & \multicolumn{1}{c}{---} & \multicolumn{1}{c}{---} & \multicolumn{1}{c}{---} & \multicolumn{1}{c}{---} & \multicolumn{1}{c}{---} & \multicolumn{1}{c}{---} & \cite{Weber1982} \\
\hline
${\rm 5S}\rightarrow {\rm 5P} \rightarrow {\rm 35D}$ & 65.4(7) & 8.6(6) & 5.44(7) & 5.83(33) & 5.81(33) & 22.7(8) & -106.0(1.1) & 20.0(8) &  This work \\ 
\hline
${\rm 5S}\rightarrow {\rm 5P} \rightarrow {\rm 37S}$ & 65.0(7) & 8.3(4) & 5.49(10) & 5.84(35) & 5.49(29) & 23.8(4) & -107.5(1.3) & 18.4(9) & This work
\\
\hline\hline
\end{tabular}
\caption{Observed collisional shift and broadening coefficients for various transitions in $^{85}$Rb.  Units of the table are in $10^{-10}$~Hz~cm$^3$.  We note that $10^{16}$~cm$^{-3}$ is the gas density of an ideal gas with pressure 0.370~mbar at 20~$^\circ$C.}
\label{tab:shifts_and_broadening}
\end{table*}

We also fit individual Lamb dips to a Lorenztian plus quadratic background to extract linewidth $\Gamma$ and frequency shift as a function of $p$.
Our values are shown in Table~\ref{tab:shifts_and_broadening} and show reasonable agreement with prior work.
All prior measurements stated in Table~\ref{tab:shifts_and_broadening} with the exception of Ref.~\cite{Sargsyan2021} were done in a similar experimental arrangement to our Doppler spectroscopy, except with gas densities $10^2$ to $10^3$ times larger than those investigated here.
Our measurement and those of Ref.~\cite{Sargsyan2021} are the only measurements we are aware of using sub-Doppler spectra.

Let us now consider the EIT spectra of Fig.~\ref{fig:raw_data} in detail, which shows a distinctly different quality.
At the lowest pressures, peaks corresponding to the $5P_{3/2}\rightarrow 35D_{3/2}$ and  $5P_{3/2}\rightarrow 35D_{5/2}$ are clearly visible for coupling laser detuning of $-300$~MHz and $0$~MHz, respectively.
Smaller peaks roughly $-75$~MHz and $-125$~MHz detuned from the large peaks are also visible; these are due to matching Doppler shifts in the probe and coupler beam with off-resonant hyperfine states of the 5P state (see Fig.~3 in Ref.~\cite{Holloway2014imaging}).
As the pressure increases, all peaks shift and broaden equally.

\begin{figure}
    \center
    \includegraphics{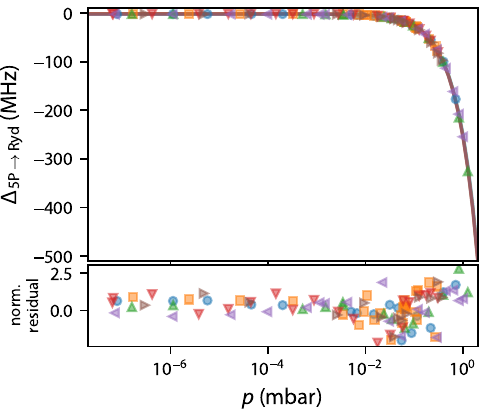}
    \caption{\label{fig:shift_example} (Top) Measured shifts of the $5{\rm S}_{1/2}\rightarrow 5{\rm P}_{3/2}\rightarrow 25{\rm D}_{5/2}$ (blue circles), $27{\rm S}_{1/2}$ (orange squares), $30{\rm D}_{5/2}$ (green upward triangles), $32{\rm S}_{1/2}$ (red downward triangles), $35{\rm D}_{5/2}$ (purple leftward triangles), $37{\rm S}_{1/2}$ (brown rightward triangles) and best fit lines (same colored curves) due to argon perturber gas.
    The best fit curves all lie on top of one another and are thus indistinguishable.  (Bottom) Fit residuals normalized to the estimated standard deviation in each measurement.}
\end{figure}

We quantify the shift and broadening by fitting the observed lines in the EIT spectra to Lorentzians to estimate the line centers and full widths at half maximum (FWHM)  at each pressure.
Known hyperfine and Doppler frequency intervals were fixed in the fits. 
The Lorentzian lineshape was chosen because it most closely matches the spectrum over all pressures; using a Gaussian lineshape increases the observed FWHMs by roughly 10~\%.
No significant change was observed in the fitted line centers between Gaussian or Lorentzian lineshapes.

The observed shift in the the EIT line, $\Delta_{\rm 5P\rightarrow Ryd}$, is shown in Fig.~\ref{fig:shift_example} as a function of $p$ for Ar perturbing gas and several different Rydberg states.
All data of this type are fit to a line.
The best fit slope determines the shift coefficient $\Delta_{\rm 5P\rightarrow Ryd}/n$; the best fit offset at $p=0$ quantifies the observed frequency offset between our reference cell and our experimental cell.
The best fit offset is $-1.65(6)$~MHz; we shall return to this shift later.

Likewise, the FWHMs as a function of $p$ are fit to lines.
The best fit slope determines the observed broadening coefficient $\Gamma_{\rm EIT}/n$; the best fit offset at $p=0$ determines the $p=0$ linewidth.
We find the linewidth at $p=0$ is 4.89(4)~MHz, which is slightly larger than the expected Doppler-mismatch broadened linewidth of 3.6~MHz~\cite{ShafferSPIE}.
The discrepancy is probably due to power broadening with the probe laser.

Table~\ref{tab:shifts_and_broadening} shows our observed shifts $\Delta_{\rm 5P\rightarrow Ryd}/n$ and broadening coefficients $\Gamma_{\rm EIT}/n$ for various EIT Rydberg transitions in $^{85}$Rb, along with comparisons to other literature values.
When comparing measurements, there are two important points to note.
First, prior experiments~\cite{Brillet1980, Weber1982, Thompson1987} measuring Rydberg transitions were typically done using two-photon excitation (of the same frequency) through a virtual P state.
As a result, the experiments required much higher Rb density, of the order of $10^{14}$~cm$^{-3}$ in order to achieve signal to noise ratios greater than one.
Thus, these experiments were typically carried out at much higher temperatures, typically around 200~$^\circ$C.

Second, because we excite through 5P, the pressure broadening $\Gamma_{\rm 5P}$ and shifts $\delta_{\rm 5P}$ of the 5P state contribute to our observed $\Gamma_{\rm EIT}$ and $\Delta_{\rm 5P\rightarrow Ryd}$.
Specifically, the probe laser, nominally resonant with $5{\rm S}\rightarrow 5{\rm P}$ when $n=0$, addresses the velocity class $k_{\rm p} v = -\delta_{\rm 5P}$, where $k_{\rm p}$ is the probe laser wavevector.
The coupling laser then becomes resonant when its detuning from resonance $\Delta_{\rm 5P\rightarrow Ryd}$ satisfies $\Delta_{\rm 5P\rightarrow Ryd} + k_{\rm c} v +\delta_{\rm 5P} = \delta$, where $k_{\rm c}$ is the wavevector of the coupling laser and $\delta$ is the collisional shift of the Rydberg state.
Thus, the observed shifts $\Delta_{\rm 5P\rightarrow Ryd}$ for the Rydberg levels shown in Table~\ref{tab:shifts_and_broadening} are related to $\delta$ via
\begin{equation}
    \label{eq:5P_shift_correction}
    \delta/n = \Delta_{\rm 5P\rightarrow Ryd}/n + \left(1-\frac{k_{\rm c}}{k_{\rm p}}\right)(\delta_{\rm 5P}/n)\,.
\end{equation}
Likewise, $\Gamma_{\rm 5P}$ will add in quadrature to the broadening of the Rydberg state $\gamma$, such that
\begin{equation}
    \label{eq:5P_broadening_correction}
    \gamma/n = \sqrt{(\Gamma_{\rm EIT}/n)^2 - (\Gamma_{\rm 5P}/n)^2}\,.
\end{equation}

The expected $\delta/n$ was first worked out by Fermi~\cite{Fermi1934} and expanded by Omont~\cite{Omont1977}.
It has two components, $\delta=\delta_{\rm sc} + \delta_{\rm p}$.
The first is the scattering of the nearly-free electron off of the perturber gas and is given in atomic units~\cite{MCWEENY1973} by
\begin{equation}
 \delta_{\rm sc}/n = 2\pi a_{\rm s}\,,
\end{equation}
where $a_{\rm s}$ is the $s$-wave scattering length of the electron off of the perturber gas.
The second contribution to the shift is due to the interaction of the atomic core with the polarizable perturber gas and is given in atomic units by 
\begin{equation}
 \delta_{\rm p}/n = -6.22(\alpha^2 \langle v\rangle )^{1/3}\,,
\end{equation}
where $\alpha$ is the polarizability of the perturber gas atom or molecule.
Note that  while $\delta_{\rm p}$ is always negative, $\delta_{\rm sc}$ (and therefore $\delta$) can be either positive or negative, depending on the sign of $a_{\rm s}$.

Likewise, Omont~\cite{Omont1977} predicted that the broadening of states with principal quantum number larger than 10 is, in atomic units,
\begin{equation}
    \gamma/n = 2\sigma_{\rm R} \bar{v}\,,
\end{equation}
where the cross section $\sigma_{\rm R}$ is again the sum of two components: the cross section due to the polarization component of the interaction potential alone, $\sigma_{\rm p}$, and the cross section due to the Fermi potential alone, $\sigma_{\rm F}$.
Expressions for $\sigma_{\rm p}$ and $\sigma_{\rm F}$ are given as (4.14)-(4.17) in Ref.~\cite{Omont1977} and shall be omitted here.

\begin{figure}
    \centering
    \includegraphics{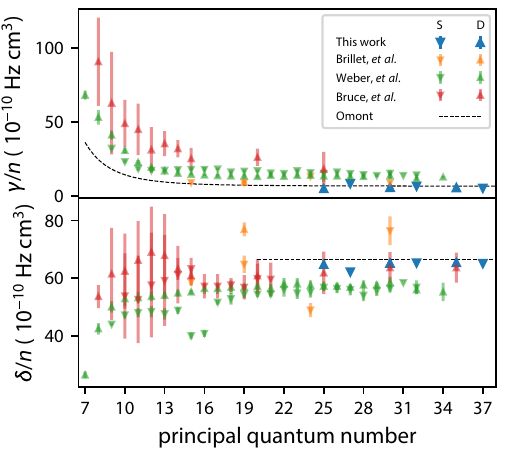}
    \caption{Measured and predicted Rb collisional broadening (top) and shifts (bottom) as a function of principal quantum number due to He perturber gas.  Our measurements are shown together with those of Brillet, {\it et al.}~\cite{Brillet1980}, Weber, {\it et al.}~\cite{Weber1982}, and Bruce, {\it et al.}~\cite{Bruce1982}, with the predictions of Omont~\cite{Omont1977}.}
    \label{fig:comparison}
\end{figure}

Figure~\ref{fig:comparison} shows our measured $\delta/n$ and $\gamma/n$ of $^{85}$Rb levels due to He perturber gas as a function of principal quantum number, together with prior measurements~\cite{Brillet1980,Weber1982,Thompson1987} and the prediction of Omont~\cite{Omont1977}.
Interestingly, our values show better agreement with Omont's predictions compared to other measurements.

\begin{table*}
\begin{tabular}{ldcddddddd}
\toprule
 & \multicolumn{2}{c}{$a_{\rm s}/a_0$} & \multicolumn{1}{c}{$\alpha/(e^2a_0^2/E_{\rm h})$} & \multicolumn{2}{c}{$\delta/n$} & \multicolumn{2}{c}{$\gamma/n$} & \multicolumn{2}{c}{$R_i$} \\
 & & & \multicolumn{1}{c}{\cite{Olney1997}} & \multicolumn{1}{c}{pred.} & \multicolumn{1}{c}{obs.} & \multicolumn{1}{c}{pred.} & \multicolumn{1}{c}{obs.} & 
 \multicolumn{1}{c}{pred.} & \multicolumn{1}{c}{obs.} \\
 & & & & \multicolumn{2}{c}{($10^{-10}$~Hz~cm$^{3}$)} & \multicolumn{2}{c}{($10^{-10}$~Hz~cm$^{3}$)} & & \\
\hline
H$_2$ & 1.61 & \cite{Randell1994} & 5.31 & 81.1 & \multicolumn{1}{c}{---} & 18 & \multicolumn{1}{c}{---} & 0.618 & \multicolumn{1}{c}{---} \\
He & 1.19 & \cite{massey1969electronic} & 1.38 & 66.6 & 64.8(7) & 6.69 & 5.8(6) & 0.466 & 0.69(4) \\
Ne & 0.24 & \cite{massey1969electronic}  & 2.66 & 6.95 & 5.1(1) & 7.95 & 5.9(3) & 0.631 & 0.24(1) \\
CH$_4$ & -2.44 & \cite{Lunt1998} & 16.5 & -175 & \multicolumn{1}{c}{---} & 28.4 & \multicolumn{1}{c}{---} & 1.13 & \multicolumn{1}{c}{---} \\
N$_2$ & 0.404 & \cite{Idziaszek2009}  & 11.5 & 5.18 & 4.1(2) & 20.1 & 23.1(6) & 1 & 1 \\
CO & 0.378 & \cite{Randell1996} & 12.6 & 2.35 & \multicolumn{1}{c}{---} & 21.3 & \multicolumn{1}{c}{---} & 1.02 & \multicolumn{1}{c}{---} \\
O$_2$ & 0.643 & \cite{Randell1994} & 10.5 & 21.2 & \multicolumn{1}{c}{---} & 18.7 & \multicolumn{1}{c}{---} & 0.994 & \multicolumn{1}{c}{---} \\
Ar & -1.7 & \cite{massey1969electronic} & 11.1 & -122 & -106.5(1.2) & 19.8 & 18.4(9) & 1.08 & 0.81(4) \\
C$_2$H$_6$ & -4.0 & \cite{Lunt1998} & 28.5 & -279 & \multicolumn{1}{c}{---} & 41.2 & \multicolumn{1}{c}{---} & 1.54 & \multicolumn{1}{c}{---} \\
CO$_2$ & -6.61 & \cite{ZIdziaszek_2008} & 16.9 & -429 & \multicolumn{1}{c}{---} & 41.8 & \multicolumn{1}{c}{---} & 1.96 & \multicolumn{1}{c}{---} \\
C$_3$H$_8$ & -4.2 & \cite{Lunt1998}  & 41 & -299 & \multicolumn{1}{c}{---} & 50.6 & \multicolumn{1}{c}{---} & 1.75 & \multicolumn{1}{c}{---} \\
Kr & -3.7 & \cite{massey1969electronic} & 16.7 & -249 & \multicolumn{1}{c}{---} & 29.4 & \multicolumn{1}{c}{---} & 1.57 & \multicolumn{1}{c}{---} \\
CF$_4$ & -2.2 & \cite{Lunt1998} & 19 & -159 & \multicolumn{1}{c}{---} & 27.1 & \multicolumn{1}{c}{---} & 1.37 & \multicolumn{1}{c}{---} \\
Xe & -6.5 & \cite{massey1969electronic} & 27.8 & -428 & \multicolumn{1}{c}{---} & 53.1 & \multicolumn{1}{c}{---} & 2.51 & \multicolumn{1}{c}{---} \\
\hline\hline
\end{tabular}
\caption{Predictions of collisional Rydberg shifts $\delta/n$ and broadening $\gamma/n$ and the ratio of Rydberg broadening to velocity changing collisions normalized to N$_2$ $R_i$ for various common contaminant or process gas species.  Predictions of $\gamma/n$ and $\delta/n$ require knowledge of the s-wave scattering length $a_{\rm s}$ of electron scattering from the perturber molecule and the perturber molecules' polarizability $\alpha$.
Note that the entries in the $a_{\rm s}$ and $\alpha$ columns are equivalent to atomic units~\cite{MCWEENY1973}, with $a_0$ being the Bohr radius, $e$ being the electric charge, and $E_{\rm h}$ being the Hartree energy.
For molecules, the value of $\alpha$ is averaged over all possible orientations.}
\label{tab:predictions}
\end{table*}

Table~\ref{tab:predictions} shows the predicted $\delta/n$ and $\gamma/n$ for high lying Rydberg states based on the theory of Omont~\cite{Omont1977} for a variety of possible contaminant gases in a vapor cell.
For all gases in Table~\ref{tab:predictions}, we have used values of $a_{\rm{s}}$ calculated by fitting very low-energy electron scattering data; $\alpha$ is taken from a calculation~\cite{Olney1997}.
While the predicted $\delta/n$ varies between roughly $-500\times10^{-10}$~Hz~cm$^3$ and $100\times10^{-10}$~Hz~cm$^3$, it is important to understand the scale of this shift coefficient.
At $T=30$~$^\circ$C and $p=10^{-6}$~mbar, an upper limit of what a vapor cell might be evacuated to prior to sealing, $n\sim 10^{10}$~cm$^{-3}$.
A hypothetical, atypically large shift coefficient of $1000\times10^{-10}$~Hz~cm$^3$ would result in $\delta \sim 1$~kHz, which is much smaller than the expected residual Doppler linewidth of 3.6~MHz~\cite{ShafferSPIE}.
The predicted $\gamma/n$ in Table~\ref{tab:predictions} cluster between $10\times10^{-10}$~Hz~cm$^3$ and $50\times10^{-10}$~Hz~cm$^3$.
At $\gamma/n = 50\times10^{10}$~Hz~cm$^3$, to get an additional broadening of 1~MHz would require $p\sim 0.01$~mbar at 30$^\circ$C.
Such pressures are unlikely in vapor cell manufacture apart from a leak.

Together with the theoretical predictions, Table~\ref{tab:predictions} also shows our observed $\delta/n$ and $\gamma/n$ for He, Ne, N$_2$, and Ar, which are determined by averaging together our results for all Rydberg transitions.
Our values for $\delta/n$ and $\gamma/n$ agree rather well with the theory of Omont, with a maximum discrepancy of about 25\,\% for Ne in both $\delta$ and $\gamma$.

\begin{figure}
    \centering
    \includegraphics{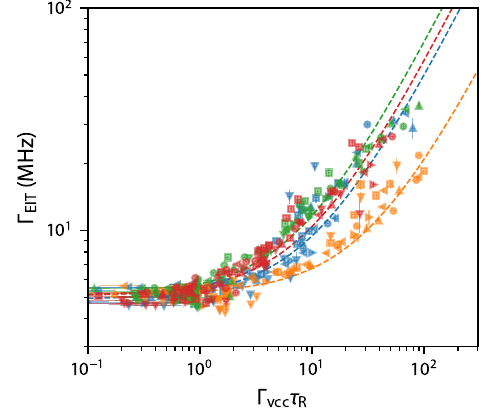}
    \caption{Measured full width at half maximum $\Gamma_{\rm EIT}$ of the Rydberg EIT lines as a function of the sub-Doppler fit parameter $\Gamma_{\rm vcc}\tau_{\rm R}$ for He (blue), Ne (orange), N$_2$ (green) and Ar (red).  Circles denote the $5{\rm S}_{1/2}\rightarrow 5{\rm P}_{3/2}\rightarrow 25{\rm D}_{5/2}$ transition; squares the $27{\rm S}_{1/2}$, upward triangles the $30{\rm D}_{5/2}$, downward triangle the $32{\rm S}_{1/2}$, leftward triangle the $35{\rm D}_{5/2}$, rightward triangle the $37{\rm S}_{1/2}$.  Dashed curves shows linear fits for each perturber gas.}
    \label{fig:EIT_linwdith_vs_Gamma_vcc_tau_R}
\end{figure}

We now return to the correlation between Rydberg EIT shifts or broadening with that of velocity-changing collisions in sub-Doppler spectra of the D$_2$ line.
Figure~\ref{fig:EIT_linwdith_vs_Gamma_vcc_tau_R} shows our observed FWHM $\Gamma_{\rm EIT}$ as a function of $\Gamma_{\rm vcc}\tau_{\rm R}$.
The data from all six Rydberg states studied are self-consistent.
We fit all the data from a given perturber gas to a line.
The slope of the best fit line gives us a measure of the ratio $\sqrt{(\gamma/n)^2 + (\Gamma_{\rm 5P}/n)^2}/(\langle \sigma v \rangle\tau_R)$.
For this discussion, we shall ignore the small contribution to the EIT linewdith from the collisional broadening of the 5P state, i.e., $\Gamma_{\rm 5P}/n\rightarrow 0$.
The $p=0$ of the best fit line again gives us a measurement of the $p=0$ linewidth of the EIT line.
Our observed ratios of $(\gamma/n)/(\langle\sigma v\rangle \tau_{\rm R})$ do not vary dramatically over the four gases measured, the total variation is only about a factor of 5.
Specifically, He, Ar, and N$_2$ all have similar values of $(\gamma/n)/(\langle\sigma v\rangle \tau_{\rm R})$; Ne is about a factor of 3-4 smaller.

Once again, we do not know $\tau_{\rm R}$, but we assume it is constant across gas species.
Thus, we define the ratio between Rydberg broadening coefficient to the rate of velocity changing collisions normalized to N$_2$, 
\begin{equation}
    R_i = \frac{[(\gamma/n)/(\langle\sigma v\rangle)]_i}{[(\gamma/n)/(\langle\sigma v\rangle)]_{\rm N_2}}\,,
\end{equation}
where $i$ represents a particular gas species.
Table~\ref{tab:predictions} shows the predicted $R_i$ for the same contaminant gas species, together with the observed quantities from the fits of Fig.~\ref{fig:EIT_linwdith_vs_Gamma_vcc_tau_R}.
To predict $R_i\propto 1/\langle \sigma v\rangle$, we use the appropriate $C_6$ coefficient, if available in the literature~\cite{Klos2023}, or scale the $C_6$ coefficient from N$_2$ using the fact that the van der Waals coefficient roughly obeys $C_6\propto \alpha$~\cite{Arpornthip2012, Eckel_2018}.
Once again, Ne stands apart as an outlier, while He, Ar, and N$_2$ show reasonable agreement.

The calculated $R_i$ shown in  Table~\ref{tab:predictions} predict that all gases   should exhibit similar ratios of Rydberg broadening to velocity changing collisions.
This is because both Rydberg broadening and VCCs are fundamentally dependent on the polarizability of the perturber atom or molecule in roughly the same way.
Specifically, $\langle \sigma v\rangle \propto \alpha^{2/5}\mu^{-3/10}$.
Likewise, the dominant contribution to the broadening of Rydberg states is due to the polarizability of the perturber atom or molecule and $\gamma_{\rm p}/n \propto \alpha^{2/3}\mu^{-1/6}$~\cite{Omont1977}.
Thus, the ratio $(\gamma/n)/\langle\sigma v\rangle\approx (\gamma_{\rm p}/n)/\langle\sigma v\rangle\propto \alpha^{4/15}\mu^{2/15}$ is extremely weakly dependent on both $\alpha$ and $\mu$.
Thus, it is not unreasonable to expect that all possible perturber gases, even if they have wildly different $\gamma/n$, will have similar ratios between Rydberg broadening and velocity changing collisions.
We thus conclude that simple measurement of VCCs in sub-Doppler spectrum could diagnose vapor cells intended for Rydberg electrometry that are suspected of having a leak, without the need for a second coupling laser to perform EIT.

\section{Conclusion}
We have revisited collisional broadening of both Rydberg states and the D$_2$ line of $^{85}$Rb to draw conclusions regarding the potential effect of contaminant gases on vapor cells manufactured for Rydberg atom electrometry.
Specifically, we investigated the changes in the Doppler, sub-Doppler, and EIT spectra as a function of He, Ne, N$_2$, and Ar perturber gas pressure.
We found that for all four of our gases, measurable effects appeared in both the sub-Doppler spectra and the EIT spectra at pressures of around 0.01~mbar.
We compared our measurements of the shifts and broadening of Rydberg EIT lines to previous measurements reported in the literature~\cite{Brillet1980,Weber1982,Thompson1987} and to the theory of Omont~\cite{Omont1977}, finding better agreement compared to previously reported results.
We found that the ratio of the broadening of Rydberg EIT lines to the rate of velocity changing collisions is weakly dependent on perturber gas species.
Finally, we generalized our results using Omont's theory and literature reported values of electron scattering lengths and polarizabilities to extend our results to multiple gases.

For the testing and manufacture of vapor cells, our work allows us to draw two conclusions:
(1) It is unlikely that contaminant gases are the sole cause of vapor cells with odd EIT lineshapes because the required pressures for measurable effects are large, on the order of 0.01~mbar, and vapor cells are generally evacuated to well below $10^{-5}$~mbar prior to sealing.
Outgassing of cell walls also is unlikely, because the total amount of trapped gases is generally well below 0.01~mbar.
Atmospheric He is known to diffuse through glass vapor cells on a time scale of years; even well-aged vapor cells containing background He at atmospheric density (roughly $1.4\times10^{14}$~cm$^{-3}$ \cite{OLIVER1984}), should exhibit insubstantial pressure broadening and minimal shifts.
Only a leak can provide the necessary amount of gas to cause a ruinous perturbation.
(2) Detecting velocity changing collisions in the sub-Doppler spectra is a sufficient test for detecting such a leak.
Such a test would be easy to implement, and consist simply of repeatedly blocking and unblocking the pump beam and quantifying the difference in the Gaussian background.
No complicated fitting of the individual Lorentzian sub-Doppler peaks would be required for a pass/fail test.

While the sub-Doppler signal may validate that the background gas pressure of a vapor cell is low enough for EIT applications, one must use caution in applying such a test to diagnose problematic vapor cells, as the root cause of defects in EIT or sub-Doppler spectra can have other potential causes.
Take, for example, our reference vapor cell, which we reported has a very consistent $-1.65(6)$~MHz offset relative to our experimental cell when $p<10^{-3}$~mbar.
In addition to this shift, the cell showed VCCs with fit parameter $\Gamma_{\rm vcc}\tau_{\rm R} = 2.7(3)$ and a extrapolated zero probe power EIT linewidth of 7.9(6)~MHz.
Taken together, these measurements seem to indicate the presence of about 100~mbar of N$_2$ gas.
First, the ratio of the measured shift of the EIT line to the FHWM of the EIT line is roughly the same as the ratio between the shift and broadening coefficients reported for N$_2$ in Table~\ref{tab:shifts_and_broadening}.
Second, the values of the EIT FWHM and $\Gamma_{\rm vcc}\tau_{\rm R}$ fall on the N$_2$ curve of Fig.~\ref{fig:EIT_linwdith_vs_Gamma_vcc_tau_R}.

To check if 100~mbar of N$_2$ gas was present in our reference cell, we constructed a vacuum chamber wherein we could destroy the cell and capture the escaping gas.
Specifically, our chamber measured the pressure rise due to the escaping gas using capacitance diaphragm and spinning rotor gauges.
After the pressure rise was quantified, the gas could be gradually released into a second, lower pressure chamber where a residual gas analyzer could measure its content.
A preliminary test was conducted by destroying a Rb cell known to be filled with 7 mbar of Ne; this test confirmed that the apparatus could measure the cell's background gas pressure and molecular content.
Upon destroying our reference cell, however, we measured that the initial pressure of the gas inside the cell was an order of magnitude lower than predicted, about $10.2\times10^{-3}$~mbar, and consisted mostly of He.
We note that if the $-1.65(6)$~MHz shift were due to He, that would correspond to a pressure of $9.4\times10^{-3}$~mbar, remarkably close to the measured value.
However, the He content would not be sufficient to explain the measured values of $\Gamma_{\rm vcc}\tau_{\rm R} = 2.7(3)$ and the zero-probe-power EIT linewidth of 7.9(6)~MHz.
Thus we return to conclusion (1) of the paper: it is unlikely that residual gases alone are the cause of irregular EIT lineshapes.
Indeed, it seems that multiple, competing effects were degrading the performance of our reference cell at the 1\,MHz level. 
The leading candidate is background electric fields due to stray charges resulting of several potential sources: (1) vapor cell manufacturing defects, (2) charge build up of ionized Rydberg atoms, and (3) cell wall interactions with alkali atoms.
Further multi-parameter studies would be required to truly understand and disentangle these competing effects, especially in the regime where residual gas pressures are below 0.1~mbar.

\section*{Acknowledgments}
The authors thank David La Mantia, Nickolas Pilgram, and Daniel Barker for technical assistance.  This work was supported by DARPA under the SAVaNT program.  The views, opinions and/or findings expressed are those of the authors and should not be interpreted as representing the official views or policies of the Department of Defense or the U.S. Government. A contribution of the U.S. government, this work is not subject to copyright in the United States.

\subsection*{Conflict of Interest}
\vspace{-3mm}
The authors have no conflicts to disclose.
\vspace{-3mm}
\subsection*{Data Availability Statement}
\vspace{-3mm}
All data presented in this review will be made available to the reader upon reasonable request to the authors.

\bibliography{main}

\end{document}